\documentclass[%
 reprint,
 superscriptaddress,
 showpacs,preprintnumbers,
 amsmath,amssymb,
 aps,
pra,
 floatfix,
]{revtex4-1}

\usepackage{graphicx}
\usepackage{dcolumn}
\usepackage{bm}
\usepackage{CJK}
\usepackage[
breaklinks,
]{hyperref}

\begin{document}
\bibliographystyle{apsrev4-1}

\title{Radiative deflection of a BaF molecular beam from the optical cycling }%

\author{Tao Chen}%

\author{Wenhao Bu}
\author{Bo Yan }%
 \email{yanbohang@zju.edu.cn}
\affiliation{%
 Department of Physics, Zhejiang University, Hangzhou, China, 310027
 }
 \affiliation{%
 Collaborative Innovation Center of Advanced Microstructures, Nanjing, China, 210093
}%


\date{\today}

\begin{abstract}

We demonstrate a quasi optical cycling for the $X(v=0)\to A(v'=0)$ transition
and a radiative force induced deflection on the buffer-gas cooled BaF
molecular beam. The laser induced fluorescence enhancement with additional
sidebands and a polarization modulation scheme indicates that the
hyperfine states and the Zeeman sublevels are closed. The quasi optical cycling
by repumping the $X(v=1)\to A(v'=0)$ leads to a $\sim$ 0.8 mm deflection of the
beam via scattering $\sim$ 150 photons per molecule, in good agreement with the
predictions from our multi-level rate equation model. Further improvement by
closing the leakage $X(v=2)$ and $\Delta$ state allows scattering thousands of
photons, and laser cooling and slowing of BaF.

\end{abstract}
\maketitle


\section{\label{section1}Introduction}

Laser cooling and trapping \cite{Chu1998} using the light scattering force
have led to lots
of fundamental breakthroughs in atomic and quantum physics, especially the
frequency standard for precision measurement \cite{Ludlow2015} and the
applications of the
degenerate quantum gases \cite{Bloch2008,Giorgini2008}. Over the last decade,
great efforts have been spared
into extending the techniques for control and cooling neutral atoms to polar
molecules \cite{Carr2009,Moses2017} due to the additional vibrational,
rotational degrees of
freedom, which provide potential novel applications in many-body physics
\cite{Wang2006,Buchler2007}, cold
controlled chemistry \cite{Krems2008,Ospelkaus2010}, and quantum simulation and
computation \cite{DeMille2002,Rabl2006,Andre2006}. While high phase
space density has been achieved in closed-shell bi-alkali molecules by external
association and adiabatic transferring techniques \cite{Ni2008,Moses2015},
producing a degenerate
open-shell molecular sample, such as alkali-alkaline-earth system, is still
under
exploration \cite{Hara2011,Pasquiou2013}. Besides,
another type of open-shell
molecule, alkaline-earth-metal monohydride and monofluoride, first proposed by
Di Rosa \cite{DiRosa2004}, can be directly laser-cooled \cite{Shuman2010}, which
has received quite great interests
in recent years.

In fact, for molecules, it is difficult to find a perfect closed optical
cycling channel to provide successive photon-molecule interactions required by
laser-cooling because of the additional complexities. Fortunately, molecules
like alkaline-earth-metal monohydride and monofluoride have special internal
level structures, leading to nearly diagonal distribution of the Franck-Condon
factors (FCFs), which results in a much simpler repumping process
\cite{DiRosa2004}, making laser cooling feasible. The earlier
experimental demonstrations of laser cooling such molecules (SrF
\cite{Shuman2009,Shuman2010} and YO
\cite{Hummon2013}) were
implemented by DeMille group and Ye group respectively, following which the
magneto-optical trapping of these two molecules was achieved short time later
\cite{Barry2014,Yeo2015,Norrgard2016}. Till
now, on one hand, the temperature of the trapped molecule samples has been
achieved lower and lower, and recently a three-dimensional molasses of
sub-Doppler temperature ($50 \mu$K) was produced for CaF molecule
\cite{Truppe2017}. On the other
hand, increasing the density of the molecular samples is urgent for further
experiments, like evaporative cooling or sympathetical cooling. Molecular
densities of $2.3\times10^5~\text{cm}^{-3}$ \cite{Truppe2017} and
$4\times10^6~\text{cm}^{-3}$ \cite{Anderegg2017} are
achieved by Hinds group and Doyle group respectively. Besides these significant
advances, laser cooling exploration on other molecules
has sprung up over the world, including YbF \cite{Tarbutt2013}, MgF
\cite{Xu2016} and BaH \cite{Iwata2017}; and now even
sub-Doppler cooling of polyatomic molecule (SrOH) has been achieved
\cite{Kozyryev2017}.

Besides the above, the heavier BaF molecule is another candidate for direct
laser cooling and trapping experiments due to the similar level structures and
the good transition wavelength ($\sim$ 900 nm) which can be easily achieved
with external cavity diode lasers \cite{Bu2016}. Besides, the effective
buffer-gas cooling of BaF to the science rovibrational levels required by laser
cooling has already been demonstrated \cite{Bu2017}. Recently, a rovibrational
cooling of a supersonic BaF beam to a rotational temperature of $\sim$ 6K with
broadband laser sources has also been reported \cite{Cournol2017}, which
provides another possible approach for preparing the cold molecular source.

In this paper, we experimentally demonstrate
the quasi-cycling transition and further observe the light scattering force
induced deflection on the buffer-gas cooled BaF molecular beam. We use
$X^2\Sigma_{1/2} \to A^2\Pi_{1/2}$ electronic transition (with the linewidth of
$\Gamma = 2\pi\times2.84$ MHz \cite{Berg1998}), which has the required highly
diagonalized FCFs, to close the vibrational branching. The cycling scheme has
been described in detail in Ref.\cite{Chen2016}. The $N=1 \to J'=1/2$ rotational
transition is employed to eliminate the rotational branching, and a sideband
modulation scheme to generate four frequency branches to cover the hyperfine
levels. In current experiment, we have not taken the leakage channel from the
$A'^2\Delta$ state into account yet. The contents are organized as following.
Section
\ref{section2} describes the experimental details. In Sec.\ref{section3}, we
present the enhancement of the laser induced fluorescence (LIF) by introducing
the $X(v=1)\to A(v'=0)$ repump laser, the sideband modulation and the
polarization modulation of the light. Furthermore, we show the deflection of the
molecular beam
induced by the quasi-cycling photon scattering. The last section gives a brief
conclusion and outlook.

\section{\label{section2}Experiment}

\begin{figure}[b]
\includegraphics[width= 0.5\textwidth]{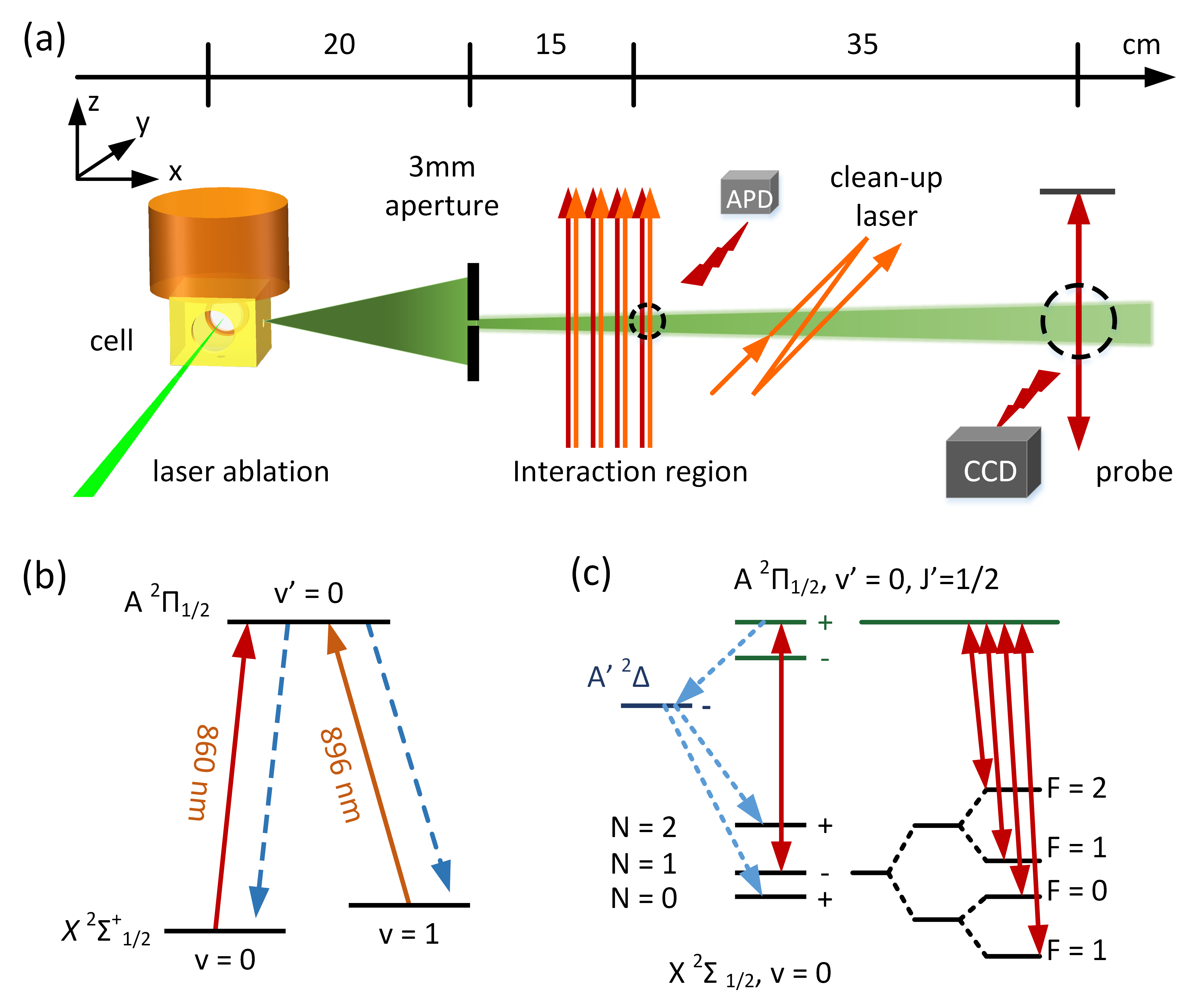}
\caption{\label{figure1}(Color online) Experimental setup and the molecular
energy levels involved. (a) Schematic
picture of the experiment. A cold BaF molecular beam along the $+\hat{x}$ direction is produced by laser ablation on a BaF$_2$
target followed by buffer-gas cooling with 4K He gas in the cell, then it
enters into the interaction region via a 3mm aperture.
The laser beams containing both pump (860nm, red) and repump (896nm, orange)
pass through the interaction region along the $+\hat{z}$
direction,and an avalanche photo diode (APD) is employed to collect the
fluorescence from the beam. The clean-up laser along the
$\hat{y}$ direction pumps the molecules from the $X(v=1)$ back into the
$X(v=0)$ ground state. Finally, the beam profiles are imaged
by a CCD camera along the $\hat{y}$ direction at 35cm downstream from the
interaction region. (b) The vibrational level structure of
BaF. A repump laser of 896nm for $X(v=1) \to A(v'=0)$ is used in our
experiment. (c) The rotational and hyperfine structures. Both the pump
and repump lasers are sideband modulated to address all four hyperfine levels of
$X(N=1)$.}
\end{figure}

Figure \ref{figure1}(a) shows the diagram of the deflection experiment. We
demonstrate the cycling scheme based on the buffer-gas cooled molecular beam of
BaF produced with the laser ablation. Different from our previous study of the
cold
collisions between BaF and He \cite{Bu2017}, the He buffer gas here flows into
the cell at a rate of 2 sccm (standard cubic centimeters per
minute). The effectively thermalized ($\sim$ 4K) mixture of He and BaF forms a beam via a 3 mm exit aperture of the cell. Another
3 mm aperture lying at 20 cm downstream from the cell filters out the molecules
with higher transverse
velocity, and collimate the beam. To deflect the molecules, we apply several
laser beams along the $\hat{z}$ direction, perpendicular to the beam
propagation. The molecule-light interaction time is controlled just by varying the pass number of the beams, and the maximum pass
number can be tuned to 8 in our experiment. The pump (860 nm) and repump (896 nm) lasers, see Fig.\ref{figure1}(b), are spatially
overlapped with a diameter of $d =$ 2 mm and powers of 160 mW and 100 mW
respectively. To
make all passes along the same direction, the laser beams are circularly
reflected around the vacuum chamber \cite{Shuman2009}. The LIF from the
$A(v'=0) \to X(v=0)$ transition is collected by an
avalanche photo diode (APD), which focuses on the first laser beam in the 10 cm
long interaction
region. The deflection probe region locates $D=$ 35 cm
away from the interaction region, and between them a clean-up laser (896 nm) with a diameter of 8 mm and power of 50 mW hits the
molecular beam to pump the molecules from the $X(v=1)$ state back to the
$X(v=0)$ state. The BaF molecular beam profiles, including the width
and position, are recorded by imaging LIF from a retroreflected laser beam (only 860 nm) on a CCD camera in probe region. The
zoom ratio of the image system is $3:1$. A band-pass filter of $860\pm10$ nm is
used to decrease the background noise from the ablation laser and other
stray lights.

To eliminate the hyperfine dark state, both the pump and repump lasers should
cover all four hyperfine levels of the $X(N=1)$ states;
see Fig.\ref{figure1}(c). Recalling the analysis in Ref.\cite{Chen2016}, a
resonant-type electro-optic modulator (EOM) with a
modulation frequency of 38 MHz is employed in our experiment, and a modulation depth of 2.6 results in the first and second
sidebands with equal amplitude, nearly matching the four hyperfine transitions
in $X(N=1,-) \to A(J'=1/2,+)$. On the other hand,
the Zeeman dark state could be remixed by applying either an angled magnetic
field or time-dependent polarization modulation \cite{Berkeland2002}.
Here we use a pockel cell to implement the polarization switching scheme and the modulation frequency is set as 1 MHz.
Additionally, both the clean-up and the deflection probe beams are sideband modulated and polarization modulated as well.

\section{\label{section3} Results and discussion}

\subsection{Quasi optical cycling}

\begin{figure}[]
\includegraphics[width= 0.4\textwidth]{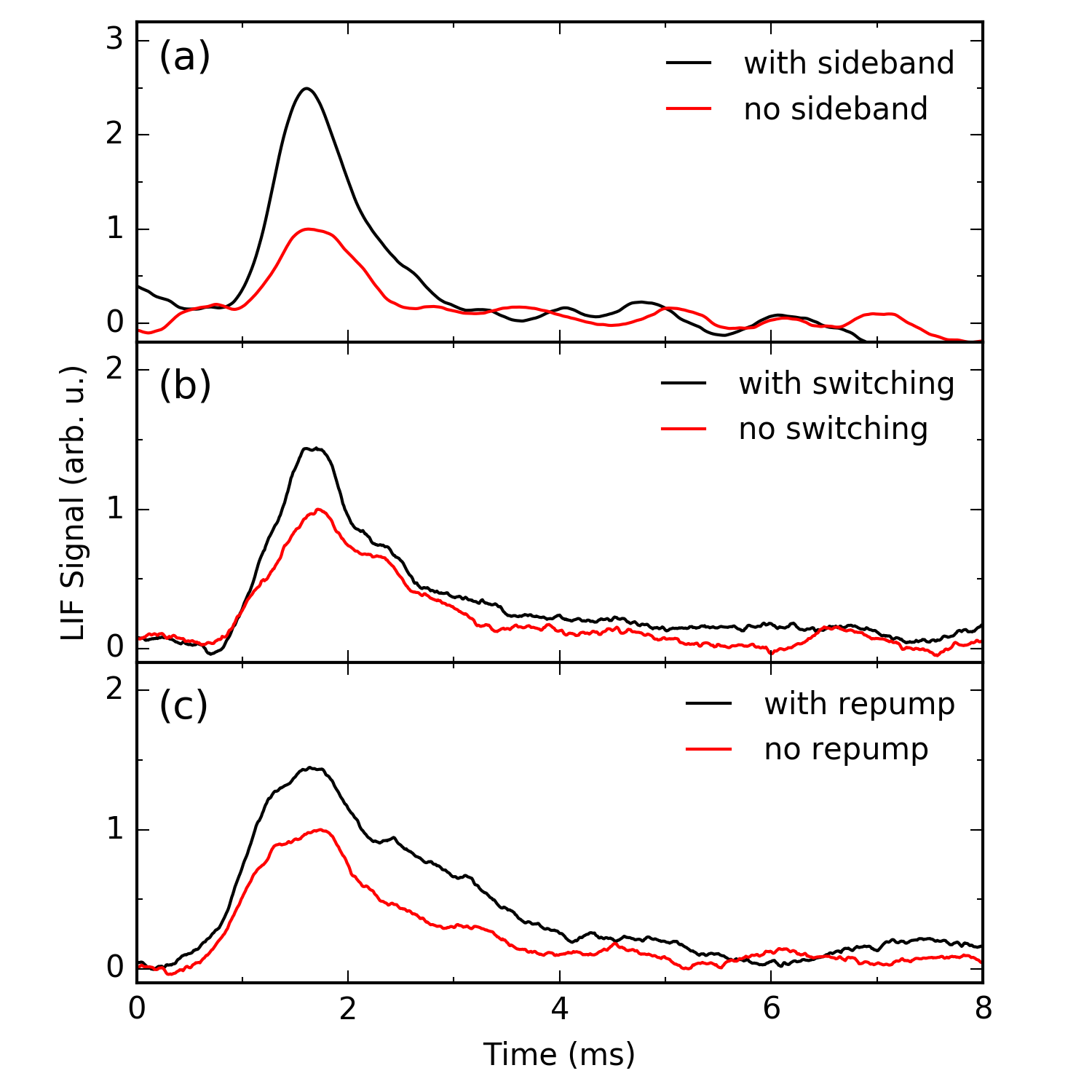}
\caption{\label{figure2}(Color online) LIF enhancement to demonstrate the
quasi optical cycling. (a) Applying sidebands to the pump laser leads to the
LIF enhancement by a factor of $\sim 2.5$. The red line indicates the APD
signal when the frequency of the pump laser hits the $F=2$ sublevel. (b) The
addition of polarization switching to both pump and repump laser to
remix Zeeman sublevels leads to $\sim 1.5\times$ enhancement. (c) The
addition of $X(v=1) \to A(v'=0)$ repump laser results in another $\sim
1.5\times$
enhancement of the LIF signal, indicating the cycling of the vibrational levels.
All the
three group signals are normalized with the peak values of the lower
signals respectively, and are averaged for hundreds of times to improve the
signal-noise ratio.}
\end{figure}

\begin{figure}[]
\includegraphics[width= 0.4\textwidth]{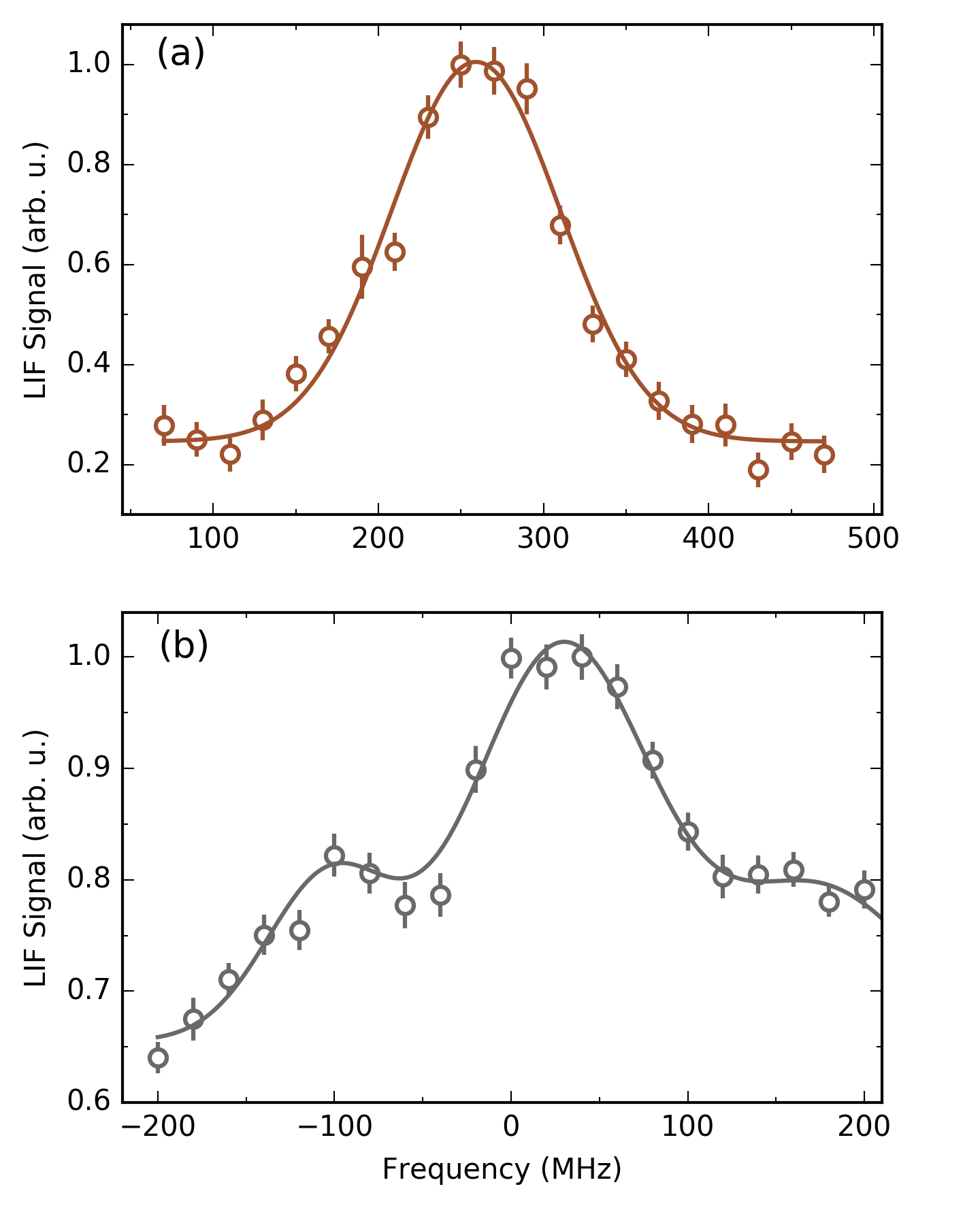}
\caption{\label{figure3}(Color online) The dependence of the LIF intensity on
the laser frequency: (a) for 860 nm pump laser; (b) for 896 nm repump laser.
The frequency scan was performed with a 38 MHz sideband modulation. For pump
laser scan, no repump laser was introduced; while for repump scan, the pump
laser was locked at the optimal point: +270 MHz. The solid lines are gaussian
fits to the data points respectively, and we lock the lasers at the
frequency for the peak: $11630.0848~ \text{cm}^{-1}$ for the 860 nm laser and
$11164.3414~ \text{cm}^{-1}$ for the 896 nm laser.}
\end{figure}

Figure \ref{figure2} shows the time of flight (ToF) LIF signals from the main
pump transition monitored by
the APD with the toggle technique applied. From Fig.\ref{figure2}(a), by
introducing the 38 MHz
sideband modulation to the pump laser to address the hyperfine sublevels, the
LIF signal is $\sim 2.5 \times$ enhanced in comparison with that when only one
single-frequency pump laser resonant with the $F=2$ sublevel applied. This can
be easily understood since much more sublevels are excited by the additional
sidebands, leading to more scattering photons before the molecules populate
the Zeeman dark states and $X(v=1)$ state. On the other hand, the ToF signal
tells us the time window of the detection. The peak LIF signal appears at
$\sim$ 1.7 ms, while the ablation
laser fires at 0 ms and the distance between the cell and the interaction
region is 35 cm, indicating the most probable velocity is $u_0 \approx$ 200
m/s. This means that the time window of the APD is about $\tau = d/u_0 = 10
~\mu\text{s}$.

The addition of the time-dependent 1 MHz polarization modulation to the pump
laser increases the LIF signal by a factor of $\sim$ 1.5; see
Fig.\ref{figure2}(b). We find that the enhancement
seems insensitive to the modulation frequency, and a 5 MHz modulation also
leads to a similar result. However, our 4+13 multi-level rate equation
(MLRE) model with the experimental parameters in Sec.\ref{section2} indicates
about $3\times$ enhancement of the scattering photon number per molecule within
$\tau = 10~\mu\text{s}$ interaction time; see
Appendix.\ref{appendix} for details. Due to the strong pump laser intensity
(the saturation factor $s$ for each sideband is $\sim$ 300),
interaction time of $10~\mu\text{s}$ is enough to pump the molecule to dark
$X(v=1)$ state, and the model shows that each molecule scatters about 18
photons, which is close to the predicted value of $N_{00} \sim
1/(1-q_{00})\approx20$ and $q_{00} = 0.9508$ \cite{Chen2016} is the FCF for the
$X(v=0)\to A(v'=0)$ transition. Consequently, the LIF enhancement with
polarization switching indicates that about 18/1.5 = 12 photons are scattered
when no switching scheme applied. This might resort to the earth's magnetic
field which can also remix the Zeeman sublevels, since from the 4+13 model we
expect only 6 photons are scattered before the molecule populates the Zeeman
dark states or $X(v=1)$ state without any remixing technique involved.

As shown in Fig.\ref{figure2}(c), the addition of the $X(v=1) \to A(v'=0)$
repump laser further makes the LIF signal $\sim 1.5 \times$ enhanced. This
indicates that the scattering photon number within $\tau = 10~\mu \text{s}$
increases to $18\times 1.5=27$, which is consistent with the predicted value
from the 4+25 MLRE model with polarization switching scheme (see
Fig.\ref{figure6} in Appendix). Till now, the quasi optical cycling has been
implemented by applying the 38 MHz sideband modulation, the 1 MHz polarization
switching scheme and the $v=1$ repump laser to close the hyperfine, the Zeeman
and the first vibrational dark states respectively. The observed LIF
enhancement agrees well with the predictions from our theoretical models.

Another important issue for the deflection experiment is the frequency of the
pump and repump lasers. Because of the different excitation rates for each
hyperfine sublevel in $X(N=1)$, we scan the frequency within several hundreds
of MHz to find an optimal position to lock the frequency of the two lasers
respectively. Figure \ref{figure3} illustrates the dependence of the LIF
signal intensity (the peak value of the ToF signal) on the laser frequency with
the 38 MHz sideband modulation. For
the pump laser, the fit tells us that the lock point should be +270 MHz,
corresponding to $11630.0848~ \text{cm}^{-1}$ (identical to the value
resolved from the in-cell spectroscopy \cite{Bu2017}); while for the repump
laser, the best point is +30 MHz, corresponding to $11164.3414~ \text{cm}^{-1}$.
The clean-up laser and probe laser in Fig.\ref{figure1} are also locked at these
two frequency points respectively \cite{Bu2016}.

\subsection{Radiative deflection}

\begin{figure}[]
\includegraphics[width= 0.45\textwidth]{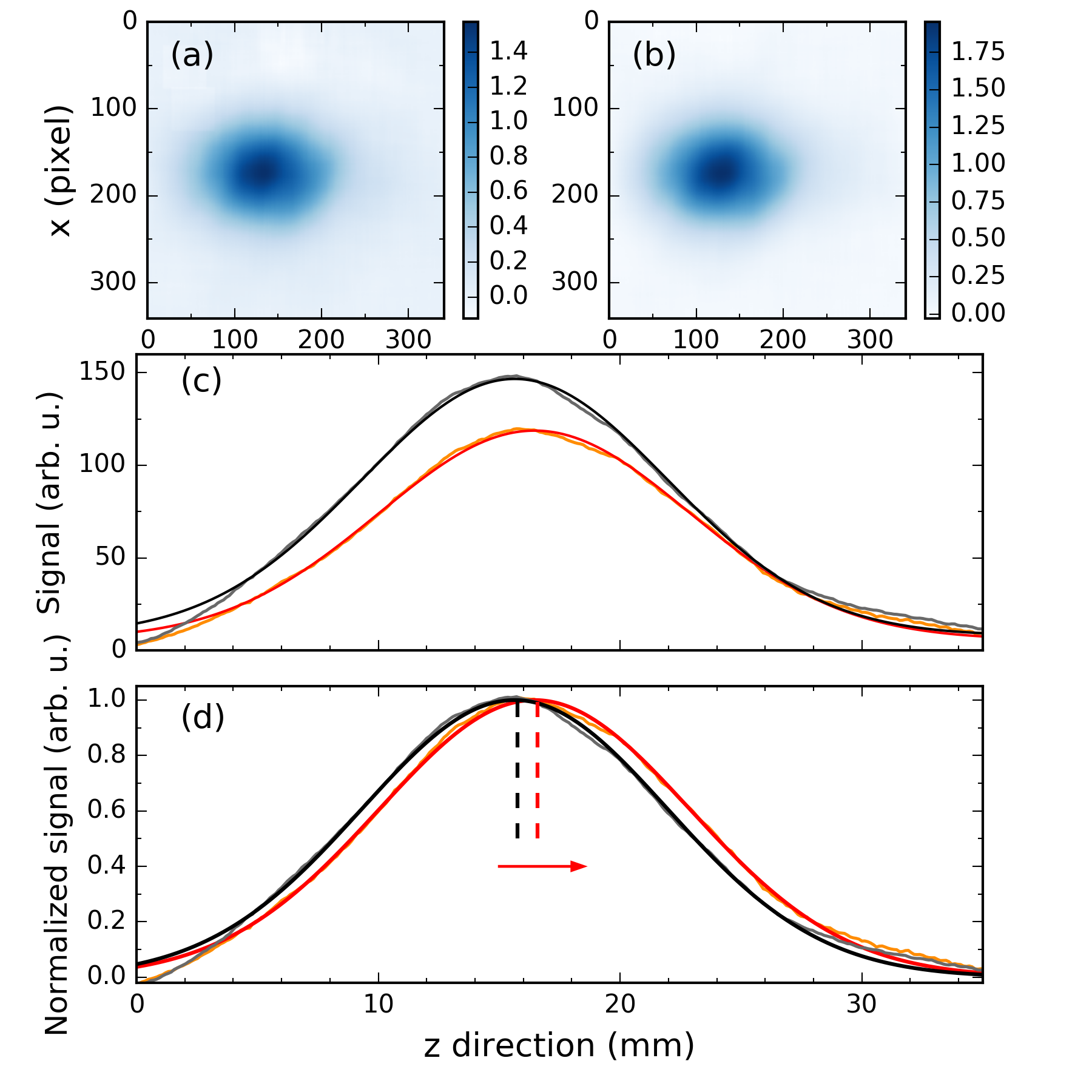}
\caption{\label{figure4}(Color online) Deflection of the BaF molecular beam
from the quasi optical cycling. (a) and (b) give the images on x-z plane of the
deflected and unperturbed molecular beam respectively. One pixel corresponds to
120 $\mu$m in reality. The $\hat{x}$ direction
reflects the width of the probe laser beam, while the $\hat{z}$ direction gives
the transverse profile of the molecular beam. (c) Integrated signal of the
image (a) and (b) along the $\hat{x}$ axis. The black and red lines are
gaussian fit to the unperturbed (gray) and deflected
(orange) signal, which gives the revival rate of 80\%. (d) Normalized plots of
the signals in (c) respectively to clearly show the deflection effect. The
width of the molecular beam remains $\sim 3$ cm with the deflection beam
applied. }
\end{figure}

The LIF enhancement for a single pass of the deflection beam in the interaction
region indicates a significant radiative force on the molecules once
the pass number $n$ increases. Figure \ref{figure4} shows the resolved molecular
beam deflection along the $\hat{z}$ direction monitored by the CCD camera for
the pass number $n=8$. The shapes of the deflected and the unperturbed beams in
the probe region are illustrated as Fig.\ref{figure4}(a) and (b) respectively.
An integration of the unperturbed image along the $\hat{x}$ axis resolves the
transverse width of the BaF molecular beam, about 3 cm, as shown in
Fig.\ref{figure4}(c). The addition of the deflection beam and clean-up
beam leads to a $\sim 0.8$ mm shift in the $+\hat{z}$ direction while the beam
width remains about 3 cm; see the normalized signals in Fig.\ref{figure4}(d).
We have also tested the effect of the $X(v=1) \to A(v'=0)$ repump laser and
clean-up laser, without which only $\sim$ 10\% molecules remain in $X(v=0)$
state after suffering the $X(v=0) \to A(v'=0)$ pump in the interaction region.
Putting the repump and clean-up laser into the system again recovers the
molecular signal to $\sim$ 80\%, which indicates effective optical pumping and
repumping. The 20\% loss is due to the leakage
$X(v\ge2)$ and the $A'{}^2\Delta$ channels \cite{Chen2016}.

Let us make an estimation of the scattering photon number $N_\text{sc}$ from
the deflection length $l$. The time required for the molecular
beam propagating from the interaction region to the probe region is
$\sim D/u_0$, then the transverse velocity changes by $\delta u = u_0l/D$.
The photon recoil momentum is given by $p = h/\lambda$, where $h$ is Planck
constant and $\lambda = 860$ nm is the wavelength of the main pump transition.
The observed deflection length $l\approx 0.8$ mm corresponds to a scattering
photon number $N_\text{sc} = m\delta u/p\approx 150 $, here $m$ is the mass of
the BaF molecule.

\begin{figure}[]
\includegraphics[width= 0.45\textwidth]{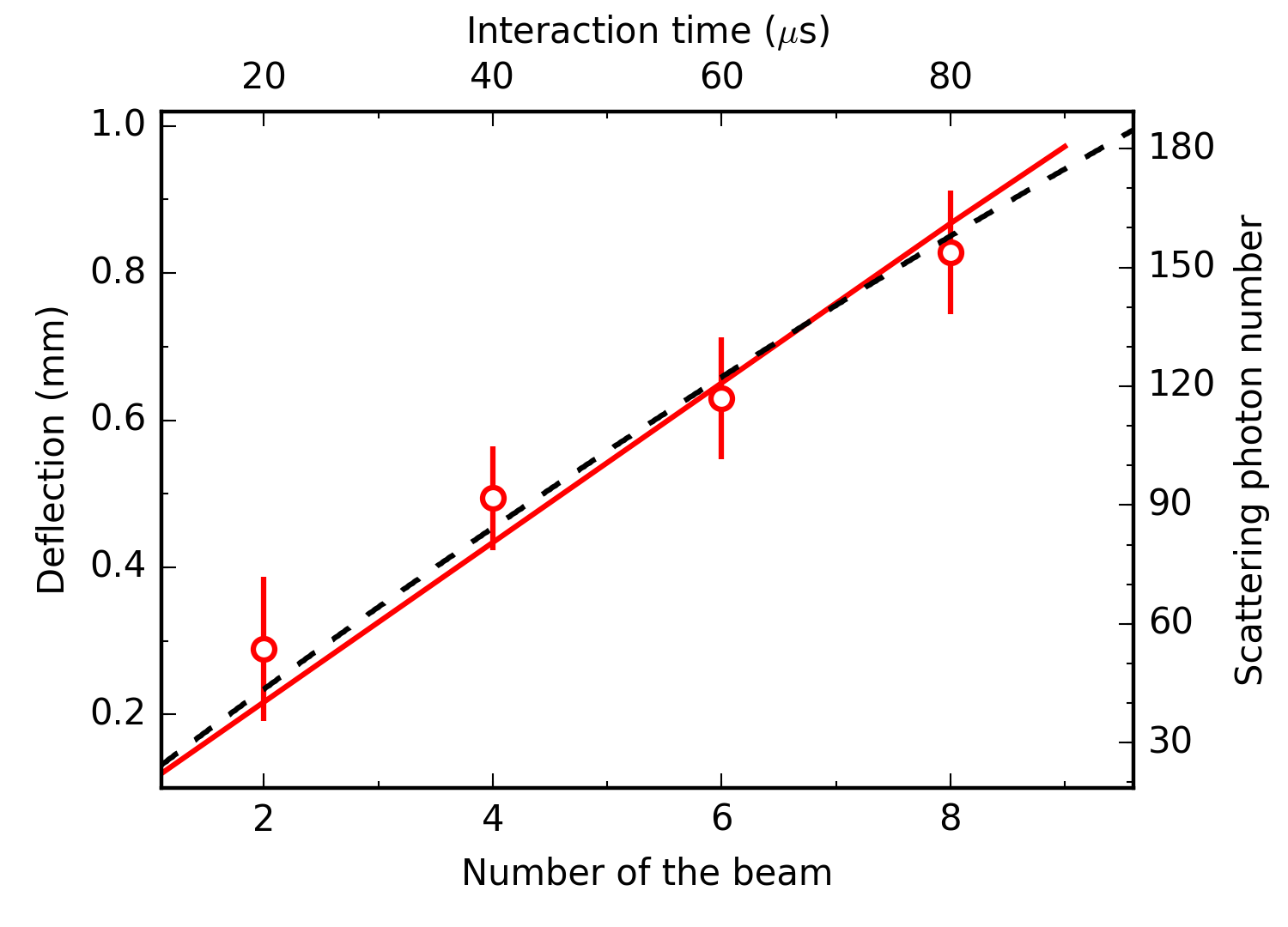}
\caption{\label{figure5}(Color online) Deflection distance as a function of the
number of the deflection beam, yielding the dependence of the scattering
photon number on the interaction time. The red solid line is a linear fit to
the measured data, illustrating that the photon scattered linearly increases
with the interaction time. The black dashed line is the numerical prediction of
the scattering from the 4+25 MLRE model with the switching scheme, which is
highly consistent with the measured data.}
\end{figure}

We have also measured the dependence of the scattered photon number
$N_\text{sc}$ on the interaction time $t=n\tau$, simply derived from deflection
length $l$ versus the pass number $n$ of the deflection beam, as shown in
Fig.\ref{figure5}. Decrease of the pass number results in a linear decrease of
the scattering photon number. The fit tells us the average scattering rate
$\Gamma_\text{sc} = 2 ~\text{MHz}$, which is only a little different from the
numerical result (also plotted in Fig.\ref{figure5}) predicted by the 4+25 MLRE
model with switching scheme. On the other hand, the theoretical maximum
scattering rate for a multi-level 4+24 system \cite{Truppe2017} is given as
$\Gamma_\text{max}= \Gamma/7 \approx 2.5 ~\text{MHz}$. The unsaturated average
scattering rate in our experiment might result from the
detunings \cite{Chen2016} of the sidebands for the hyperfine
transitions.

\section{\label{section4} Conclusion}

To summarize, we have clearly shown the evidence of the quasi optical cycling
and further the radiative force from the $\sim 156$ scattering photons with
only one additional $X(v=1)\to A(v'=0)$ repump. By applying the 38 MHz sideband
modulation to the pump and repump lasers, the hyperfine dark states are
eliminated. For Zeeman dark states, we have employed the 1 MHz polarization
switching scheme to remix them to the cycling. Putting all these techniques
together and increasing the pass number of the beam to achieve longer
interaction time, we
have observed a significant transverse deflection ($\sim 0.8$ mm) of the BaF
molecular beam, indicating a scattering rate of $\sim 2$ MHz, which agrees well
with the theoretical prediction from our MLRE model.

By adjusting the detunings of the pump and repump
lasers, retroreflecting the both laser beams and providing sufficient
interaction length, the molecular beam should be transversely cooled.
Furthermore, the scattering photon number required for loading the beam
to a trap is about $mu_0/p \approx 6.5\times 10^4$ with a frequency-chirped or
white light to longitudinal slow the beam, as a consequence, another
transition, for example, $X(v=0) \to B(v''=0)$, might be employed to improve
the
scattering rate \cite{Truppe2017a}. To build a magneto-optical
trapping (MOT) of BaF, the addition of $X(v=2) \to A(v'=1)$ repump laser should
be required due to the calculated larger branching ratio of $\sim 1.5\times
10^{-3}$ for $A(v'=0) \to X(v=2)$ \cite{Chen2016} than those of CaF
\cite{Zhelyazkova2014} and SrF \cite{Shuman2009}. Besides the RF-MOT with
polarization switching \cite{Anderegg2017}, our previously proposed microwave
mediated MOT ($\mu$-MOT) \cite{Xie2016} might be another candidate for our
future laser cooling and trapping experiment.

\begin{acknowledgments}

We acknowledge the support from the National Natural Science Foundation of China
under Grant No. 91636104, Zhejiang Provincial Natural Science Foundation under
Grant No. LZ18A040001, the Fundamental Research Funds for the Central
Universities Grant No. 2016QNA3007. We thank Yong Xia for useful discussions.
\end{acknowledgments}

\appendix
\section{\label{appendix} Multi-level rate equation model}

The rate equations to describe the time evolution of the populated fraction in
each sublevel for a multi-level system is given as \cite{Metcalf1999}
\begin{eqnarray}
 \frac{dN_l}{dt} &=& \Gamma\sum\limits_u r_{l,u}N_u +
\sum\limits_{u,p}R_{l,u,p}(N_u-N_l), \nonumber \\[0.3em]
 \frac{dN_u}{dt} &=& -\Gamma N_u + \sum\limits_{l,p} R_{l,u,p} (N_l-N_u),
\end{eqnarray}
where $N_l$ and $N_u$ are the populated fractions for the $l$-th sublevel in the
ground state and the $u$-th sublevel in the  excited state, $\Gamma$ is
the spontaneous decay rate of the excited state, $r_{l,u}$ is
the branching ratio for $u \to l$ transitions (see the values in
Ref.\cite{Chen2016}). $R_{l,u,p} =
\frac{\Gamma}{2}\frac{r_{l,u}s_p}{1+(2\Delta_p/\Gamma)^2}$ is the excitation
rate for $l \to u$ transition from the $p$-th laser beam \cite{Tarbutt2015}, and
$s_p$ is the
saturation factor and $\Delta_p$ is the detuning. For the evaluation of
$R_{l,u,p}$ with polarization switching scheme, we should take the selection
rules into account, i.e., $R_{l,u,p}=0$ for $m_u = m_l + 1$ when $\sigma_p =
\sigma_-$ and $m_u = m_l - 1$ when $\sigma_p = \sigma_+$. The scattered photon
number at time $t_0$ is evaluated from $N_\text{sc}(t_0) =
\sum\limits_u\int_0^{t_0}q_{00}\Gamma N_u dt$.

\begin{figure}[b]
\includegraphics[width= 0.45\textwidth]{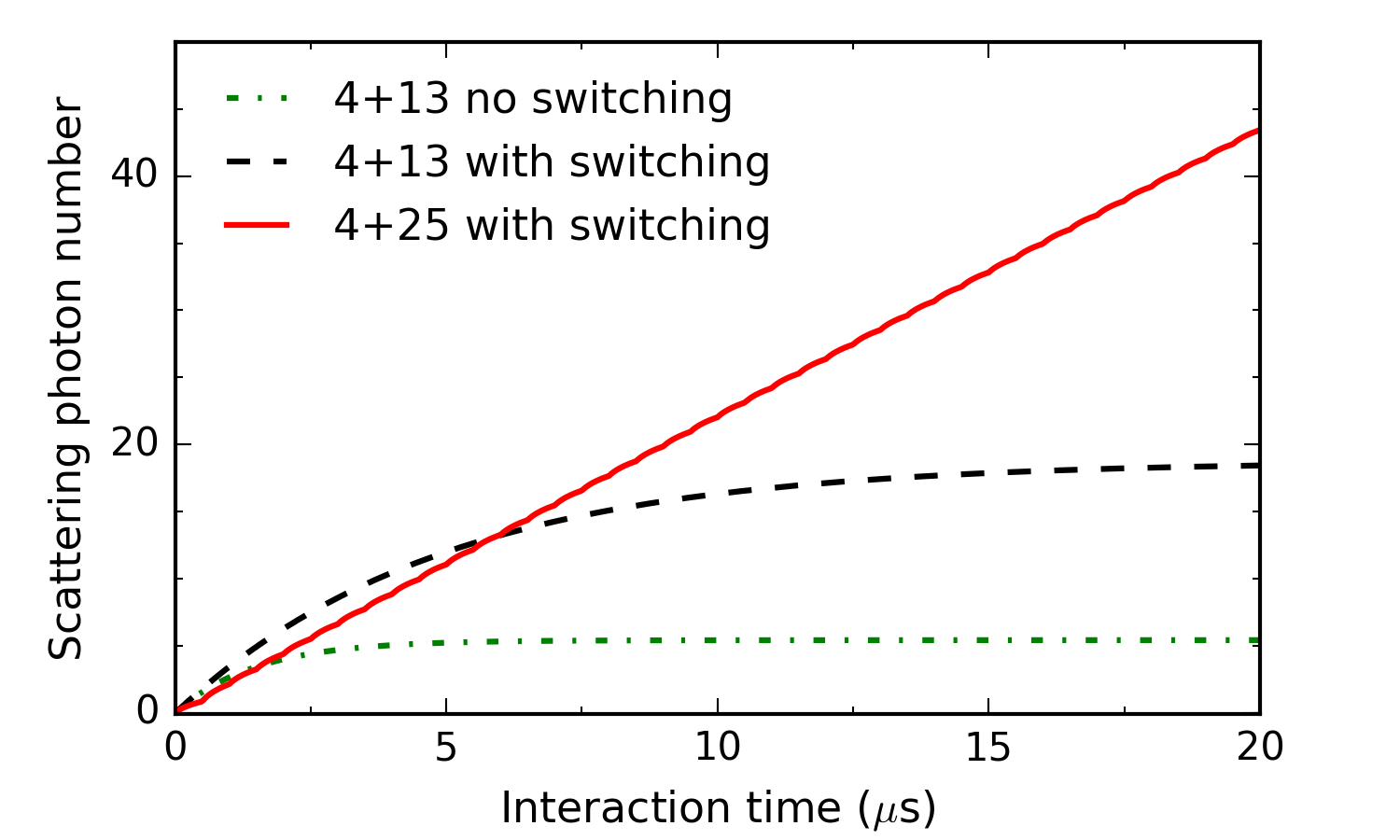}
\caption{\label{figure6}(Color online) Predicted scattering photon number as a
function of the interaction time within 20 $\mu$s. The values are rapidly
saturated for 4+13 model, even with polarization switching scheme to eliminate
the Zeeman dark state. With both switching scheme and $X(v=1)$ repump, the 4+25
model shows a linear relation between scattering photon number and interaction
time, but the number saturates to $\sim$ 600 with a rather longer interaction
time (not plotted in the figure).}
\end{figure}

We firstly build a 4+13 model with linearly polarized laser applied,
considering 4 excited states in $A(v'=0,J'=1/2,+)$ and 12 sublevels in
$X(v=0,N=1)$. The 13rd level is the assumed loss channel with a branching ratio
of $q_\text{loss} = 1- q_{00} = 0.05$. For linearly polarized excitation, the
$X(v=0,N=1,F=2,m_F=\pm 2)$ sublevels are dark states. Our numerical calculation
indicates that the molecule will loss to the 13rd level or populate the Zeeman
dark states just after scattering $\sim 6$ photons; see Fig.\ref{figure6}. By
introducing the 1 MHz polarization modulation, the model shows that scattered
photon number increases to $\sim 20$, three times larger than that without
switching, before the molecule entirely populates the dark states.

To close the loss channel, we add the $X(v=1)\to A(v'=0)$ repump laser to our
model, i.e., 4+25 model. Besides the 4 excited states and 12 sublevels for
$X(v=0,N=1)$ and $X(v=1,N=1)$ respectively, the other loss channels, for
example, $X(v\ge 2)$ and $A'^2\Delta$ states, are all labeled as the 25-th
level with a total branching ratio $q_\text{loss}= 1-q_{00}-q_{01} = 1.6\times
10^{-3}$. This model indicates that the molecule suffers from
nearly successive photon scattering within 100 $\mu$s (larger than the
interaction time in our deflection experiment), and finally the scattering
process terminates after $\sim$ 600 photons (close to the value of
$1/q_\text{loss}$) are scattered for an interaction time of about 1 ms. Figure
\ref{figure6} shows the scattering photon number as a function of the
interaction time within 20 $\mu$s. For 10 $\mu$s interaction time, the addition
of the $X(v=1)$ repump laser only increases the scattering number by a factor
of $\sim 1.5$, which is consistent with our experimental observation. Finally,
to achieve laser cooling of BaF, additions of $X(v=2)$ repump laser and
microwave
remixing of $\Delta \to X(N=0,2)$ channels \cite{Chen2016} are necessary to
scatter thousands of photons.

\bibliography{cycling_and_deflection}

\begin{thebibliography}{41}%
\makeatletter
\providecommand \@ifxundefined [1]{%
 \@ifx{#1\undefined}
}%
\providecommand \@ifnum [1]{%
 \ifnum #1\expandafter \@firstoftwo
 \else \expandafter \@secondoftwo
 \fi
}%
\providecommand \@ifx [1]{%
 \ifx #1\expandafter \@firstoftwo
 \else \expandafter \@secondoftwo
 \fi
}%
\providecommand \natexlab [1]{#1}%
\providecommand \enquote  [1]{``#1''}%
\providecommand \bibnamefont  [1]{#1}%
\providecommand \bibfnamefont [1]{#1}%
\providecommand \citenamefont [1]{#1}%
\providecommand \href@noop [0]{\@secondoftwo}%
\providecommand \href [0]{\begingroup \@sanitize@url \@href}%
\providecommand \@href[1]{\@@startlink{#1}\@@href}%
\providecommand \@@href[1]{\endgroup#1\@@endlink}%
\providecommand \@sanitize@url [0]{\catcode `\\12\catcode `\$12\catcode
  `\&12\catcode `\#12\catcode `\^12\catcode `\_12\catcode `\%12\relax}%
\providecommand \@@startlink[1]{}%
\providecommand \@@endlink[0]{}%
\providecommand \url  [0]{\begingroup\@sanitize@url \@url }%
\providecommand \@url [1]{\endgroup\@href {#1}{\urlprefix }}%
\providecommand \urlprefix  [0]{URL }%
\providecommand \Eprint [0]{\href }%
\providecommand \doibase [0]{http://dx.doi.org/}%
\providecommand \selectlanguage [0]{\@gobble}%
\providecommand \bibinfo  [0]{\@secondoftwo}%
\providecommand \bibfield  [0]{\@secondoftwo}%
\providecommand \translation [1]{[#1]}%
\providecommand \BibitemOpen [0]{}%
\providecommand \bibitemStop [0]{}%
\providecommand \bibitemNoStop [0]{.\EOS\space}%
\providecommand \EOS [0]{\spacefactor3000\relax}%
\providecommand \BibitemShut  [1]{\csname bibitem#1\endcsname}%
\let\auto@bib@innerbib\@empty
\bibitem [{\citenamefont {Chu}(1998)}]{Chu1998}%
  \BibitemOpen
  \bibfield  {author} {\bibinfo {author} {\bibfnamefont {S.}~\bibnamefont
  {Chu}},\ }\href {\doibase 10.1103/RevModPhys.70.685} {\bibfield  {journal}
  {\bibinfo  {journal} {Rev. Mod. Phys.}\ }\textbf {\bibinfo {volume} {70}},\
  \bibinfo {pages} {685} (\bibinfo {year} {1998})}\BibitemShut {NoStop}%
\bibitem [{\citenamefont {Ludlow}\ \emph {et~al.}(2015)\citenamefont {Ludlow},
  \citenamefont {Boyd}, \citenamefont {Ye}, \citenamefont {Peik},\ and\
  \citenamefont {Schmidt}}]{Ludlow2015}%
  \BibitemOpen
  \bibfield  {author} {\bibinfo {author} {\bibfnamefont {A.~D.}\ \bibnamefont
  {Ludlow}}, \bibinfo {author} {\bibfnamefont {M.~M.}\ \bibnamefont {Boyd}},
  \bibinfo {author} {\bibfnamefont {J.}~\bibnamefont {Ye}}, \bibinfo {author}
  {\bibfnamefont {E.}~\bibnamefont {Peik}}, \ and\ \bibinfo {author}
  {\bibfnamefont {P.~O.}\ \bibnamefont {Schmidt}},\ }\href {\doibase
  10.1103/RevModPhys.87.637} {\bibfield  {journal} {\bibinfo  {journal} {Rev.
  Mod. Phys.}\ }\textbf {\bibinfo {volume} {87}},\ \bibinfo {pages} {637}
  (\bibinfo {year} {2015})}\BibitemShut {NoStop}%
\bibitem [{\citenamefont {Bloch}\ \emph {et~al.}(2008)\citenamefont {Bloch},
  \citenamefont {Dalibard},\ and\ \citenamefont {Zwerger}}]{Bloch2008}%
  \BibitemOpen
  \bibfield  {author} {\bibinfo {author} {\bibfnamefont {I.}~\bibnamefont
  {Bloch}}, \bibinfo {author} {\bibfnamefont {J.}~\bibnamefont {Dalibard}}, \
  and\ \bibinfo {author} {\bibfnamefont {W.}~\bibnamefont {Zwerger}},\ }\href
  {\doibase 10.1103/RevModPhys.80.885} {\bibfield  {journal} {\bibinfo
  {journal} {Rev. Mod. Phys.}\ }\textbf {\bibinfo {volume} {80}},\ \bibinfo
  {pages} {885} (\bibinfo {year} {2008})}\BibitemShut {NoStop}%
\bibitem [{\citenamefont {Giorgini}\ \emph {et~al.}(2008)\citenamefont
  {Giorgini}, \citenamefont {Pitaevskii},\ and\ \citenamefont
  {Stringari}}]{Giorgini2008}%
  \BibitemOpen
  \bibfield  {author} {\bibinfo {author} {\bibfnamefont {S.}~\bibnamefont
  {Giorgini}}, \bibinfo {author} {\bibfnamefont {L.~P.}\ \bibnamefont
  {Pitaevskii}}, \ and\ \bibinfo {author} {\bibfnamefont {S.}~\bibnamefont
  {Stringari}},\ }\href {\doibase 10.1103/RevModPhys.80.1215} {\bibfield
  {journal} {\bibinfo  {journal} {Rev. Mod. Phys.}\ }\textbf {\bibinfo {volume}
  {80}},\ \bibinfo {pages} {1215} (\bibinfo {year} {2008})}\BibitemShut
  {NoStop}%
\bibitem [{\citenamefont {Carr}\ \emph {et~al.}(2009)\citenamefont {Carr},
  \citenamefont {DeMille}, \citenamefont {Krems},\ and\ \citenamefont
  {Ye}}]{Carr2009}%
  \BibitemOpen
  \bibfield  {author} {\bibinfo {author} {\bibfnamefont {L.~D.}\ \bibnamefont
  {Carr}}, \bibinfo {author} {\bibfnamefont {D.}~\bibnamefont {DeMille}},
  \bibinfo {author} {\bibfnamefont {R.~V.}\ \bibnamefont {Krems}}, \ and\
  \bibinfo {author} {\bibfnamefont {J.}~\bibnamefont {Ye}},\ }\href
  {http://stacks.iop.org/1367-2630/11/i=5/a=055049} {\bibfield  {journal}
  {\bibinfo  {journal} {New Journal of Physics}\ }\textbf {\bibinfo {volume}
  {11}},\ \bibinfo {pages} {055049} (\bibinfo {year} {2009})}\BibitemShut
  {NoStop}%
\bibitem [{\citenamefont {Moses}\ \emph {et~al.}(2017)\citenamefont {Moses},
  \citenamefont {Covey}, \citenamefont {Miecnikowski}, \citenamefont {Jin},\
  and\ \citenamefont {Ye}}]{Moses2017}%
  \BibitemOpen
  \bibfield  {author} {\bibinfo {author} {\bibfnamefont {S.~A.}\ \bibnamefont
  {Moses}}, \bibinfo {author} {\bibfnamefont {J.~P.}\ \bibnamefont {Covey}},
  \bibinfo {author} {\bibfnamefont {M.~T.}\ \bibnamefont {Miecnikowski}},
  \bibinfo {author} {\bibfnamefont {D.~S.}\ \bibnamefont {Jin}}, \ and\
  \bibinfo {author} {\bibfnamefont {J.}~\bibnamefont {Ye}},\ }\href {\doibase
  http://dx.doi.org/10.1038/nphys3985} {\bibfield  {journal} {\bibinfo
  {journal} {Nat. Phys.}\ }\textbf {\bibinfo {volume} {13}},\ \bibinfo {pages}
  {13} (\bibinfo {year} {2017})}\BibitemShut {NoStop}%
\bibitem [{\citenamefont {Wang}\ \emph {et~al.}(2006)\citenamefont {Wang},
  \citenamefont {Lukin},\ and\ \citenamefont {Demler}}]{Wang2006}%
  \BibitemOpen
  \bibfield  {author} {\bibinfo {author} {\bibfnamefont {D.-W.}\ \bibnamefont
  {Wang}}, \bibinfo {author} {\bibfnamefont {M.~D.}\ \bibnamefont {Lukin}}, \
  and\ \bibinfo {author} {\bibfnamefont {E.}~\bibnamefont {Demler}},\ }\href
  {\doibase 10.1103/PhysRevLett.97.180413} {\bibfield  {journal} {\bibinfo
  {journal} {Phys. Rev. Lett.}\ }\textbf {\bibinfo {volume} {97}},\ \bibinfo
  {pages} {180413} (\bibinfo {year} {2006})}\BibitemShut {NoStop}%
\bibitem [{\citenamefont {Buchler}\ \emph {et~al.}(2007)\citenamefont
  {Buchler}, \citenamefont {Micheli},\ and\ \citenamefont
  {Zoller}}]{Buchler2007}%
  \BibitemOpen
  \bibfield  {author} {\bibinfo {author} {\bibfnamefont {H.~P.}\ \bibnamefont
  {Buchler}}, \bibinfo {author} {\bibfnamefont {A.}~\bibnamefont {Micheli}}, \
  and\ \bibinfo {author} {\bibfnamefont {P.}~\bibnamefont {Zoller}},\ }\href
  {\doibase http://dx.doi.org/10.1038/nphys678} {\bibfield  {journal} {\bibinfo
   {journal} {Nat. Phys.}\ }\textbf {\bibinfo {volume} {3}},\ \bibinfo {pages}
  {726} (\bibinfo {year} {2007})}\BibitemShut {NoStop}%
\bibitem [{\citenamefont {Krems}(2008)}]{Krems2008}%
  \BibitemOpen
  \bibfield  {author} {\bibinfo {author} {\bibfnamefont {R.~V.}\ \bibnamefont
  {Krems}},\ }\href {\doibase 10.1039/B802322K} {\bibfield  {journal} {\bibinfo
   {journal} {Phys. Chem. Chem. Phys.}\ }\textbf {\bibinfo {volume} {10}},\
  \bibinfo {pages} {4079} (\bibinfo {year} {2008})}\BibitemShut {NoStop}%
\bibitem [{\citenamefont {Ospelkaus}\ \emph {et~al.}(2010)\citenamefont
  {Ospelkaus}, \citenamefont {Ni}, \citenamefont {Wang}, \citenamefont
  {de~Miranda}, \citenamefont {Neyenhuis}, \citenamefont {Qu\'em\'ener},
  \citenamefont {Julienne}, \citenamefont {Bohn}, \citenamefont {Jin},\ and\
  \citenamefont {Ye}}]{Ospelkaus2010}%
  \BibitemOpen
  \bibfield  {author} {\bibinfo {author} {\bibfnamefont {S.}~\bibnamefont
  {Ospelkaus}}, \bibinfo {author} {\bibfnamefont {K.-K.}\ \bibnamefont {Ni}},
  \bibinfo {author} {\bibfnamefont {D.}~\bibnamefont {Wang}}, \bibinfo {author}
  {\bibfnamefont {M.~H.~G.}\ \bibnamefont {de~Miranda}}, \bibinfo {author}
  {\bibfnamefont {B.}~\bibnamefont {Neyenhuis}}, \bibinfo {author}
  {\bibfnamefont {G.}~\bibnamefont {Qu\'em\'ener}}, \bibinfo {author}
  {\bibfnamefont {P.~S.}\ \bibnamefont {Julienne}}, \bibinfo {author}
  {\bibfnamefont {J.~L.}\ \bibnamefont {Bohn}}, \bibinfo {author}
  {\bibfnamefont {D.~S.}\ \bibnamefont {Jin}}, \ and\ \bibinfo {author}
  {\bibfnamefont {J.}~\bibnamefont {Ye}},\ }\href {\doibase
  10.1126/science.1184121} {\bibfield  {journal} {\bibinfo  {journal}
  {Science}\ }\textbf {\bibinfo {volume} {327}},\ \bibinfo {pages} {853}
  (\bibinfo {year} {2010})}\BibitemShut {NoStop}%
\bibitem [{\citenamefont {DeMille}(2002)}]{DeMille2002}%
  \BibitemOpen
  \bibfield  {author} {\bibinfo {author} {\bibfnamefont {D.}~\bibnamefont
  {DeMille}},\ }\href {\doibase 10.1103/PhysRevLett.88.067901} {\bibfield
  {journal} {\bibinfo  {journal} {Phys. Rev. Lett.}\ }\textbf {\bibinfo
  {volume} {88}},\ \bibinfo {pages} {067901} (\bibinfo {year}
  {2002})}\BibitemShut {NoStop}%
\bibitem [{\citenamefont {Rabl}\ \emph {et~al.}(2006)\citenamefont {Rabl},
  \citenamefont {DeMille}, \citenamefont {Doyle}, \citenamefont {Lukin},
  \citenamefont {Schoelkopf},\ and\ \citenamefont {Zoller}}]{Rabl2006}%
  \BibitemOpen
  \bibfield  {author} {\bibinfo {author} {\bibfnamefont {P.}~\bibnamefont
  {Rabl}}, \bibinfo {author} {\bibfnamefont {D.}~\bibnamefont {DeMille}},
  \bibinfo {author} {\bibfnamefont {J.~M.}\ \bibnamefont {Doyle}}, \bibinfo
  {author} {\bibfnamefont {M.~D.}\ \bibnamefont {Lukin}}, \bibinfo {author}
  {\bibfnamefont {R.~J.}\ \bibnamefont {Schoelkopf}}, \ and\ \bibinfo {author}
  {\bibfnamefont {P.}~\bibnamefont {Zoller}},\ }\href {\doibase
  10.1103/PhysRevLett.97.033003} {\bibfield  {journal} {\bibinfo  {journal}
  {Phys. Rev. Lett.}\ }\textbf {\bibinfo {volume} {97}},\ \bibinfo {pages}
  {033003} (\bibinfo {year} {2006})}\BibitemShut {NoStop}%
\bibitem [{\citenamefont {Andre}\ \emph {et~al.}(2006)\citenamefont {Andre},
  \citenamefont {DeMille}, \citenamefont {Doyle}, \citenamefont {Lukin},
  \citenamefont {Maxwell}, \citenamefont {Rabl}, \citenamefont {Schoelkopf},\
  and\ \citenamefont {Zoller}}]{Andre2006}%
  \BibitemOpen
  \bibfield  {author} {\bibinfo {author} {\bibfnamefont {A.}~\bibnamefont
  {Andre}}, \bibinfo {author} {\bibfnamefont {D.}~\bibnamefont {DeMille}},
  \bibinfo {author} {\bibfnamefont {J.~M.}\ \bibnamefont {Doyle}}, \bibinfo
  {author} {\bibfnamefont {M.~D.}\ \bibnamefont {Lukin}}, \bibinfo {author}
  {\bibfnamefont {S.~E.}\ \bibnamefont {Maxwell}}, \bibinfo {author}
  {\bibfnamefont {P.}~\bibnamefont {Rabl}}, \bibinfo {author} {\bibfnamefont
  {R.~J.}\ \bibnamefont {Schoelkopf}}, \ and\ \bibinfo {author} {\bibfnamefont
  {P.}~\bibnamefont {Zoller}},\ }\href {\doibase
  http://dx.doi.org/10.1038/nphys386} {\bibfield  {journal} {\bibinfo
  {journal} {Nat. Phys.}\ }\textbf {\bibinfo {volume} {2}},\ \bibinfo {pages}
  {636} (\bibinfo {year} {2006})}\BibitemShut {NoStop}%
\bibitem [{\citenamefont {Ni}\ \emph {et~al.}(2008)\citenamefont {Ni},
  \citenamefont {Ospelkaus}, \citenamefont {de~Miranda}, \citenamefont {Pe'er},
  \citenamefont {Neyenhuis}, \citenamefont {Zirbel}, \citenamefont
  {Kotochigova}, \citenamefont {Julienne}, \citenamefont {Jin},\ and\
  \citenamefont {Ye}}]{Ni2008}%
  \BibitemOpen
  \bibfield  {author} {\bibinfo {author} {\bibfnamefont {K.-K.}\ \bibnamefont
  {Ni}}, \bibinfo {author} {\bibfnamefont {S.}~\bibnamefont {Ospelkaus}},
  \bibinfo {author} {\bibfnamefont {M.~H.~G.}\ \bibnamefont {de~Miranda}},
  \bibinfo {author} {\bibfnamefont {A.}~\bibnamefont {Pe'er}}, \bibinfo
  {author} {\bibfnamefont {B.}~\bibnamefont {Neyenhuis}}, \bibinfo {author}
  {\bibfnamefont {J.~J.}\ \bibnamefont {Zirbel}}, \bibinfo {author}
  {\bibfnamefont {S.}~\bibnamefont {Kotochigova}}, \bibinfo {author}
  {\bibfnamefont {P.~S.}\ \bibnamefont {Julienne}}, \bibinfo {author}
  {\bibfnamefont {D.~S.}\ \bibnamefont {Jin}}, \ and\ \bibinfo {author}
  {\bibfnamefont {J.}~\bibnamefont {Ye}},\ }\href {\doibase
  10.1126/science.1163861} {\bibfield  {journal} {\bibinfo  {journal}
  {Science}\ }\textbf {\bibinfo {volume} {322}},\ \bibinfo {pages} {231}
  (\bibinfo {year} {2008})}\BibitemShut {NoStop}%
\bibitem [{\citenamefont {Moses}\ \emph {et~al.}(2015)\citenamefont {Moses},
  \citenamefont {Covey}, \citenamefont {Miecnikowski}, \citenamefont {Yan},
  \citenamefont {Gadway}, \citenamefont {Ye},\ and\ \citenamefont
  {Jin}}]{Moses2015}%
  \BibitemOpen
  \bibfield  {author} {\bibinfo {author} {\bibfnamefont {S.~A.}\ \bibnamefont
  {Moses}}, \bibinfo {author} {\bibfnamefont {J.~P.}\ \bibnamefont {Covey}},
  \bibinfo {author} {\bibfnamefont {M.~T.}\ \bibnamefont {Miecnikowski}},
  \bibinfo {author} {\bibfnamefont {B.}~\bibnamefont {Yan}}, \bibinfo {author}
  {\bibfnamefont {B.}~\bibnamefont {Gadway}}, \bibinfo {author} {\bibfnamefont
  {J.}~\bibnamefont {Ye}}, \ and\ \bibinfo {author} {\bibfnamefont {D.~S.}\
  \bibnamefont {Jin}},\ }\href {\doibase 10.1126/science.aac6400} {\bibfield
  {journal} {\bibinfo  {journal} {Science}\ }\textbf {\bibinfo {volume}
  {350}},\ \bibinfo {pages} {659} (\bibinfo {year} {2015})}\BibitemShut
  {NoStop}%
\bibitem [{\citenamefont {Hara}\ \emph {et~al.}(2011)\citenamefont {Hara},
  \citenamefont {Takasu}, \citenamefont {Yamaoka}, \citenamefont {Doyle},\ and\
  \citenamefont {Takahashi}}]{Hara2011}%
  \BibitemOpen
  \bibfield  {author} {\bibinfo {author} {\bibfnamefont {H.}~\bibnamefont
  {Hara}}, \bibinfo {author} {\bibfnamefont {Y.}~\bibnamefont {Takasu}},
  \bibinfo {author} {\bibfnamefont {Y.}~\bibnamefont {Yamaoka}}, \bibinfo
  {author} {\bibfnamefont {J.~M.}\ \bibnamefont {Doyle}}, \ and\ \bibinfo
  {author} {\bibfnamefont {Y.}~\bibnamefont {Takahashi}},\ }\href {\doibase
  10.1103/PhysRevLett.106.205304} {\bibfield  {journal} {\bibinfo  {journal}
  {Phys. Rev. Lett.}\ }\textbf {\bibinfo {volume} {106}},\ \bibinfo {pages}
  {205304} (\bibinfo {year} {2011})}\BibitemShut {NoStop}%
\bibitem [{\citenamefont {Pasquiou}\ \emph {et~al.}(2013)\citenamefont
  {Pasquiou}, \citenamefont {Bayerle}, \citenamefont {Tzanova}, \citenamefont
  {Stellmer}, \citenamefont {Szczepkowski}, \citenamefont {Parigger},
  \citenamefont {Grimm},\ and\ \citenamefont {Schreck}}]{Pasquiou2013}%
  \BibitemOpen
  \bibfield  {author} {\bibinfo {author} {\bibfnamefont {B.}~\bibnamefont
  {Pasquiou}}, \bibinfo {author} {\bibfnamefont {A.}~\bibnamefont {Bayerle}},
  \bibinfo {author} {\bibfnamefont {S.~M.}\ \bibnamefont {Tzanova}}, \bibinfo
  {author} {\bibfnamefont {S.}~\bibnamefont {Stellmer}}, \bibinfo {author}
  {\bibfnamefont {J.}~\bibnamefont {Szczepkowski}}, \bibinfo {author}
  {\bibfnamefont {M.}~\bibnamefont {Parigger}}, \bibinfo {author}
  {\bibfnamefont {R.}~\bibnamefont {Grimm}}, \ and\ \bibinfo {author}
  {\bibfnamefont {F.}~\bibnamefont {Schreck}},\ }\href {\doibase
  10.1103/PhysRevA.88.023601} {\bibfield  {journal} {\bibinfo  {journal} {Phys.
  Rev. A}\ }\textbf {\bibinfo {volume} {88}},\ \bibinfo {pages} {023601}
  (\bibinfo {year} {2013})}\BibitemShut {NoStop}%
\bibitem [{\citenamefont {Di~Rosa}(2004)}]{DiRosa2004}%
  \BibitemOpen
  \bibfield  {author} {\bibinfo {author} {\bibfnamefont {M.~D.}\ \bibnamefont
  {Di~Rosa}},\ }\href {\doibase 10.1140/epjd/e2004-00167-2} {\bibfield
  {journal} {\bibinfo  {journal} {The European Physical Journal D - Atomic,
  Molecular, Optical and Plasma Physics}\ }\textbf {\bibinfo {volume} {31}},\
  \bibinfo {pages} {395} (\bibinfo {year} {2004})}\BibitemShut {NoStop}%
\bibitem [{\citenamefont {Shuman}\ \emph {et~al.}(2010)\citenamefont {Shuman},
  \citenamefont {Barry},\ and\ \citenamefont {DeMille}}]{Shuman2010}%
  \BibitemOpen
  \bibfield  {author} {\bibinfo {author} {\bibfnamefont {E.~S.}\ \bibnamefont
  {Shuman}}, \bibinfo {author} {\bibfnamefont {J.~F.}\ \bibnamefont {Barry}}, \
  and\ \bibinfo {author} {\bibfnamefont {D.}~\bibnamefont {DeMille}},\ }\href
  {\doibase 10.1038/nature09443} {\bibfield  {journal} {\bibinfo  {journal}
  {Nature}\ }\textbf {\bibinfo {volume} {467}},\ \bibinfo {pages} {820}
  (\bibinfo {year} {2010})}\BibitemShut {NoStop}%
\bibitem [{\citenamefont {Shuman}\ \emph {et~al.}(2009)\citenamefont {Shuman},
  \citenamefont {Barry}, \citenamefont {Glenn},\ and\ \citenamefont
  {DeMille}}]{Shuman2009}%
  \BibitemOpen
  \bibfield  {author} {\bibinfo {author} {\bibfnamefont {E.~S.}\ \bibnamefont
  {Shuman}}, \bibinfo {author} {\bibfnamefont {J.~F.}\ \bibnamefont {Barry}},
  \bibinfo {author} {\bibfnamefont {D.~R.}\ \bibnamefont {Glenn}}, \ and\
  \bibinfo {author} {\bibfnamefont {D.}~\bibnamefont {DeMille}},\ }\href
  {\doibase 10.1103/PhysRevLett.103.223001} {\bibfield  {journal} {\bibinfo
  {journal} {Phys. Rev. Lett.}\ }\textbf {\bibinfo {volume} {103}},\ \bibinfo
  {pages} {223001} (\bibinfo {year} {2009})}\BibitemShut {NoStop}%
\bibitem [{\citenamefont {Hummon}\ \emph {et~al.}(2013)\citenamefont {Hummon},
  \citenamefont {Yeo}, \citenamefont {Stuhl}, \citenamefont {Collopy},
  \citenamefont {Xia},\ and\ \citenamefont {Ye}}]{Hummon2013}%
  \BibitemOpen
  \bibfield  {author} {\bibinfo {author} {\bibfnamefont {M.~T.}\ \bibnamefont
  {Hummon}}, \bibinfo {author} {\bibfnamefont {M.}~\bibnamefont {Yeo}},
  \bibinfo {author} {\bibfnamefont {B.~K.}\ \bibnamefont {Stuhl}}, \bibinfo
  {author} {\bibfnamefont {A.~L.}\ \bibnamefont {Collopy}}, \bibinfo {author}
  {\bibfnamefont {Y.}~\bibnamefont {Xia}}, \ and\ \bibinfo {author}
  {\bibfnamefont {J.}~\bibnamefont {Ye}},\ }\href {\doibase
  10.1103/PhysRevLett.110.143001} {\bibfield  {journal} {\bibinfo  {journal}
  {Phys. Rev. Lett.}\ }\textbf {\bibinfo {volume} {110}},\ \bibinfo {pages}
  {143001} (\bibinfo {year} {2013})}\BibitemShut {NoStop}%
\bibitem [{\citenamefont {Barry}\ \emph {et~al.}(2014)\citenamefont {Barry},
  \citenamefont {McCarron}, \citenamefont {Norrgard}, \citenamefont
  {Steinecker},\ and\ \citenamefont {DeMille}}]{Barry2014}%
  \BibitemOpen
  \bibfield  {author} {\bibinfo {author} {\bibfnamefont {J.~F.}\ \bibnamefont
  {Barry}}, \bibinfo {author} {\bibfnamefont {D.~J.}\ \bibnamefont {McCarron}},
  \bibinfo {author} {\bibfnamefont {E.~B.}\ \bibnamefont {Norrgard}}, \bibinfo
  {author} {\bibfnamefont {M.~H.}\ \bibnamefont {Steinecker}}, \ and\ \bibinfo
  {author} {\bibfnamefont {D.}~\bibnamefont {DeMille}},\ }\href
  {http://www.nature.com/nature/journal/v512/n7514/full/nature13634.html}
  {\bibfield  {journal} {\bibinfo  {journal} {Nature}\ }\textbf {\bibinfo
  {volume} {512}},\ \bibinfo {pages} {286} (\bibinfo {year}
  {2014})}\BibitemShut {NoStop}%
\bibitem [{\citenamefont {Yeo}\ \emph {et~al.}(2015)\citenamefont {Yeo},
  \citenamefont {Hummon}, \citenamefont {Collopy}, \citenamefont {Yan},
  \citenamefont {Hemmerling}, \citenamefont {Chae}, \citenamefont {Doyle},\
  and\ \citenamefont {Ye}}]{Yeo2015}%
  \BibitemOpen
  \bibfield  {author} {\bibinfo {author} {\bibfnamefont {M.}~\bibnamefont
  {Yeo}}, \bibinfo {author} {\bibfnamefont {M.~T.}\ \bibnamefont {Hummon}},
  \bibinfo {author} {\bibfnamefont {A.~L.}\ \bibnamefont {Collopy}}, \bibinfo
  {author} {\bibfnamefont {B.}~\bibnamefont {Yan}}, \bibinfo {author}
  {\bibfnamefont {B.}~\bibnamefont {Hemmerling}}, \bibinfo {author}
  {\bibfnamefont {E.}~\bibnamefont {Chae}}, \bibinfo {author} {\bibfnamefont
  {J.~M.}\ \bibnamefont {Doyle}}, \ and\ \bibinfo {author} {\bibfnamefont
  {J.}~\bibnamefont {Ye}},\ }\href {\doibase 10.1103/PhysRevLett.114.223003}
  {\bibfield  {journal} {\bibinfo  {journal} {Phys. Rev. Lett.}\ }\textbf
  {\bibinfo {volume} {114}},\ \bibinfo {pages} {223003} (\bibinfo {year}
  {2015})}\BibitemShut {NoStop}%
\bibitem [{\citenamefont {Norrgard}\ \emph {et~al.}(2016)\citenamefont
  {Norrgard}, \citenamefont {McCarron}, \citenamefont {Steinecker},
  \citenamefont {Tarbutt},\ and\ \citenamefont {DeMille}}]{Norrgard2016}%
  \BibitemOpen
  \bibfield  {author} {\bibinfo {author} {\bibfnamefont {E.~B.}\ \bibnamefont
  {Norrgard}}, \bibinfo {author} {\bibfnamefont {D.~J.}\ \bibnamefont
  {McCarron}}, \bibinfo {author} {\bibfnamefont {M.~H.}\ \bibnamefont
  {Steinecker}}, \bibinfo {author} {\bibfnamefont {M.~R.}\ \bibnamefont
  {Tarbutt}}, \ and\ \bibinfo {author} {\bibfnamefont {D.}~\bibnamefont
  {DeMille}},\ }\href {\doibase 10.1103/PhysRevLett.116.063004} {\bibfield
  {journal} {\bibinfo  {journal} {Phys. Rev. Lett.}\ }\textbf {\bibinfo
  {volume} {116}},\ \bibinfo {pages} {063004} (\bibinfo {year}
  {2016})}\BibitemShut {NoStop}%
\bibitem [{\citenamefont {Truppe}\ \emph
  {et~al.}(2017{\natexlab{a}})\citenamefont {Truppe}, \citenamefont {Williams},
  \citenamefont {Hambach}, \citenamefont {Caldwell}, \citenamefont {Fitch},
  \citenamefont {Hinds}, \citenamefont {Sauer},\ and\ \citenamefont
  {Tarbutt}}]{Truppe2017}%
  \BibitemOpen
  \bibfield  {author} {\bibinfo {author} {\bibfnamefont {S.}~\bibnamefont
  {Truppe}}, \bibinfo {author} {\bibfnamefont {H.~J.}\ \bibnamefont
  {Williams}}, \bibinfo {author} {\bibfnamefont {M.}~\bibnamefont {Hambach}},
  \bibinfo {author} {\bibfnamefont {L.}~\bibnamefont {Caldwell}}, \bibinfo
  {author} {\bibfnamefont {N.~J.}\ \bibnamefont {Fitch}}, \bibinfo {author}
  {\bibfnamefont {E.~A.}\ \bibnamefont {Hinds}}, \bibinfo {author}
  {\bibfnamefont {B.~E.}\ \bibnamefont {Sauer}}, \ and\ \bibinfo {author}
  {\bibfnamefont {M.~R.}\ \bibnamefont {Tarbutt}},\ }\href
  {http://dx.doi.org/10.1038/nphys4241} {\bibfield  {journal} {\bibinfo
  {journal} {Nat. Phys.}\ }\textbf {\bibinfo {volume} {advance online
  publication}} (\bibinfo {year} {2017}{\natexlab{a}})}\BibitemShut {NoStop}%
\bibitem [{\citenamefont {Anderegg}\ \emph {et~al.}(2017)\citenamefont
  {Anderegg}, \citenamefont {Augenbraun}, \citenamefont {Chae}, \citenamefont
  {Hemmerling}, \citenamefont {Hutzler}, \citenamefont {Ravi}, \citenamefont
  {Collopy}, \citenamefont {Ye}, \citenamefont {Ketterle},\ and\ \citenamefont
  {Doyle}}]{Anderegg2017}%
  \BibitemOpen
  \bibfield  {author} {\bibinfo {author} {\bibfnamefont {L.}~\bibnamefont
  {Anderegg}}, \bibinfo {author} {\bibfnamefont {B.~L.}\ \bibnamefont
  {Augenbraun}}, \bibinfo {author} {\bibfnamefont {E.}~\bibnamefont {Chae}},
  \bibinfo {author} {\bibfnamefont {B.}~\bibnamefont {Hemmerling}}, \bibinfo
  {author} {\bibfnamefont {N.~R.}\ \bibnamefont {Hutzler}}, \bibinfo {author}
  {\bibfnamefont {A.}~\bibnamefont {Ravi}}, \bibinfo {author} {\bibfnamefont
  {A.}~\bibnamefont {Collopy}}, \bibinfo {author} {\bibfnamefont
  {J.}~\bibnamefont {Ye}}, \bibinfo {author} {\bibfnamefont {W.}~\bibnamefont
  {Ketterle}}, \ and\ \bibinfo {author} {\bibfnamefont {J.~M.}\ \bibnamefont
  {Doyle}},\ }\href {\doibase 10.1103/PhysRevLett.119.103201} {\bibfield
  {journal} {\bibinfo  {journal} {Phys. Rev. Lett.}\ }\textbf {\bibinfo
  {volume} {119}},\ \bibinfo {pages} {103201} (\bibinfo {year}
  {2017})}\BibitemShut {NoStop}%
\bibitem [{\citenamefont {Tarbutt}\ \emph {et~al.}(2013)\citenamefont
  {Tarbutt}, \citenamefont {Sauer}, \citenamefont {Hudson},\ and\ \citenamefont
  {Hinds}}]{Tarbutt2013}%
  \BibitemOpen
  \bibfield  {author} {\bibinfo {author} {\bibfnamefont {M.~R.}\ \bibnamefont
  {Tarbutt}}, \bibinfo {author} {\bibfnamefont {B.~E.}\ \bibnamefont {Sauer}},
  \bibinfo {author} {\bibfnamefont {J.~J.}\ \bibnamefont {Hudson}}, \ and\
  \bibinfo {author} {\bibfnamefont {E.~A.}\ \bibnamefont {Hinds}},\ }\href
  {http://stacks.iop.org/1367-2630/15/i=5/a=053034} {\bibfield  {journal}
  {\bibinfo  {journal} {New Journal of Physics}\ }\textbf {\bibinfo {volume}
  {15}},\ \bibinfo {pages} {053034} (\bibinfo {year} {2013})}\BibitemShut
  {NoStop}%
\bibitem [{\citenamefont {Xu}\ \emph {et~al.}(2016)\citenamefont {Xu},
  \citenamefont {Yin}, \citenamefont {Wei}, \citenamefont {Xia},\ and\
  \citenamefont {Yin}}]{Xu2016}%
  \BibitemOpen
  \bibfield  {author} {\bibinfo {author} {\bibfnamefont {L.}~\bibnamefont
  {Xu}}, \bibinfo {author} {\bibfnamefont {Y.}~\bibnamefont {Yin}}, \bibinfo
  {author} {\bibfnamefont {B.}~\bibnamefont {Wei}}, \bibinfo {author}
  {\bibfnamefont {Y.}~\bibnamefont {Xia}}, \ and\ \bibinfo {author}
  {\bibfnamefont {J.}~\bibnamefont {Yin}},\ }\href {\doibase
  10.1103/PhysRevA.93.013408} {\bibfield  {journal} {\bibinfo  {journal} {Phys.
  Rev. A}\ }\textbf {\bibinfo {volume} {93}},\ \bibinfo {pages} {013408}
  (\bibinfo {year} {2016})}\BibitemShut {NoStop}%
\bibitem [{\citenamefont {Iwata}\ \emph {et~al.}(2017)\citenamefont {Iwata},
  \citenamefont {McNally},\ and\ \citenamefont {Zelevinsky}}]{Iwata2017}%
  \BibitemOpen
  \bibfield  {author} {\bibinfo {author} {\bibfnamefont {G.~Z.}\ \bibnamefont
  {Iwata}}, \bibinfo {author} {\bibfnamefont {R.~L.}\ \bibnamefont {McNally}},
  \ and\ \bibinfo {author} {\bibfnamefont {T.}~\bibnamefont {Zelevinsky}},\
  }\href {\doibase 10.1103/PhysRevA.96.022509} {\bibfield  {journal} {\bibinfo
  {journal} {Phys. Rev. A}\ }\textbf {\bibinfo {volume} {96}},\ \bibinfo
  {pages} {022509} (\bibinfo {year} {2017})}\BibitemShut {NoStop}%
\bibitem [{\citenamefont {Kozyryev}\ \emph {et~al.}(2017)\citenamefont
  {Kozyryev}, \citenamefont {Baum}, \citenamefont {Matsuda}, \citenamefont
  {Augenbraun}, \citenamefont {Anderegg}, \citenamefont {Sedlack},\ and\
  \citenamefont {Doyle}}]{Kozyryev2017}%
  \BibitemOpen
  \bibfield  {author} {\bibinfo {author} {\bibfnamefont {I.}~\bibnamefont
  {Kozyryev}}, \bibinfo {author} {\bibfnamefont {L.}~\bibnamefont {Baum}},
  \bibinfo {author} {\bibfnamefont {K.}~\bibnamefont {Matsuda}}, \bibinfo
  {author} {\bibfnamefont {B.~L.}\ \bibnamefont {Augenbraun}}, \bibinfo
  {author} {\bibfnamefont {L.}~\bibnamefont {Anderegg}}, \bibinfo {author}
  {\bibfnamefont {A.~P.}\ \bibnamefont {Sedlack}}, \ and\ \bibinfo {author}
  {\bibfnamefont {J.~M.}\ \bibnamefont {Doyle}},\ }\href {\doibase
  10.1103/PhysRevLett.118.173201} {\bibfield  {journal} {\bibinfo  {journal}
  {Phys. Rev. Lett.}\ }\textbf {\bibinfo {volume} {118}},\ \bibinfo {pages}
  {173201} (\bibinfo {year} {2017})}\BibitemShut {NoStop}%
\bibitem [{\citenamefont {Bu}\ \emph {et~al.}(2016)\citenamefont {Bu},
  \citenamefont {Liu}, \citenamefont {Xie},\ and\ \citenamefont
  {Yan}}]{Bu2016}%
  \BibitemOpen
  \bibfield  {author} {\bibinfo {author} {\bibfnamefont {W.}~\bibnamefont
  {Bu}}, \bibinfo {author} {\bibfnamefont {M.}~\bibnamefont {Liu}}, \bibinfo
  {author} {\bibfnamefont {D.}~\bibnamefont {Xie}}, \ and\ \bibinfo {author}
  {\bibfnamefont {B.}~\bibnamefont {Yan}},\ }\href
  {http://dx.doi.org/10.1063/1.4963361} {\bibfield  {journal} {\bibinfo
  {journal} {Review of Scientific Instruments}\ }\textbf {\bibinfo {volume}
  {87}},\ \bibinfo {pages} {096102} (\bibinfo {year} {2016})}\BibitemShut
  {NoStop}%
\bibitem [{\citenamefont {Bu}\ \emph {et~al.}(2017)\citenamefont {Bu},
  \citenamefont {Chen}, \citenamefont {Lv},\ and\ \citenamefont
  {Yan}}]{Bu2017}%
  \BibitemOpen
  \bibfield  {author} {\bibinfo {author} {\bibfnamefont {W.}~\bibnamefont
  {Bu}}, \bibinfo {author} {\bibfnamefont {T.}~\bibnamefont {Chen}}, \bibinfo
  {author} {\bibfnamefont {G.}~\bibnamefont {Lv}}, \ and\ \bibinfo {author}
  {\bibfnamefont {B.}~\bibnamefont {Yan}},\ }\href {\doibase
  10.1103/PhysRevA.95.032701} {\bibfield  {journal} {\bibinfo  {journal} {Phys.
  Rev. A}\ }\textbf {\bibinfo {volume} {95}},\ \bibinfo {pages} {032701}
  (\bibinfo {year} {2017})}\BibitemShut {NoStop}%
\bibitem [{\citenamefont {Cournol}\ \emph {et~al.}(2017)\citenamefont
  {Cournol}, \citenamefont {Pillet}, \citenamefont {Lignier},\ and\
  \citenamefont {Comparat}}]{Cournol2017}%
  \BibitemOpen
  \bibfield  {author} {\bibinfo {author} {\bibfnamefont {A.}~\bibnamefont
  {Cournol}}, \bibinfo {author} {\bibfnamefont {P.}~\bibnamefont {Pillet}},
  \bibinfo {author} {\bibfnamefont {H.}~\bibnamefont {Lignier}}, \ and\
  \bibinfo {author} {\bibfnamefont {D.}~\bibnamefont {Comparat}},\ }\href
  {https://arxiv.org/abs/1709.06797} {\bibfield  {journal} {\bibinfo  {journal}
  {arXiv}\ }\textbf {\bibinfo {volume} {1709.06797}} (\bibinfo {year}
  {2017})}\BibitemShut {NoStop}%
\bibitem [{\citenamefont {Berg}\ \emph {et~al.}(1998)\citenamefont {Berg},
  \citenamefont {Gador}, \citenamefont {Husain}, \citenamefont {Ludwigs},\ and\
  \citenamefont {Royen}}]{Berg1998}%
  \BibitemOpen
  \bibfield  {author} {\bibinfo {author} {\bibfnamefont {L.-E.}\ \bibnamefont
  {Berg}}, \bibinfo {author} {\bibfnamefont {N.}~\bibnamefont {Gador}},
  \bibinfo {author} {\bibfnamefont {D.}~\bibnamefont {Husain}}, \bibinfo
  {author} {\bibfnamefont {H.}~\bibnamefont {Ludwigs}}, \ and\ \bibinfo
  {author} {\bibfnamefont {P.}~\bibnamefont {Royen}},\ }\href {\doibase
  http://dx.doi.org/10.1016/S0009-2614(98)00149-3} {\bibfield  {journal}
  {\bibinfo  {journal} {Chem. Phys. Lett.}\ }\textbf {\bibinfo {volume}
  {287}},\ \bibinfo {pages} {89 } (\bibinfo {year} {1998})}\BibitemShut
  {NoStop}%
\bibitem [{\citenamefont {Chen}\ \emph {et~al.}(2016)\citenamefont {Chen},
  \citenamefont {Bu},\ and\ \citenamefont {Yan}}]{Chen2016}%
  \BibitemOpen
  \bibfield  {author} {\bibinfo {author} {\bibfnamefont {T.}~\bibnamefont
  {Chen}}, \bibinfo {author} {\bibfnamefont {W.}~\bibnamefont {Bu}}, \ and\
  \bibinfo {author} {\bibfnamefont {B.}~\bibnamefont {Yan}},\ }\href {\doibase
  10.1103/PhysRevA.94.063415} {\bibfield  {journal} {\bibinfo  {journal} {Phys.
  Rev. A}\ }\textbf {\bibinfo {volume} {94}},\ \bibinfo {pages} {063415}
  (\bibinfo {year} {2016})}\BibitemShut {NoStop}%
\bibitem [{\citenamefont {Berkeland}\ and\ \citenamefont
  {Boshier}(2002)}]{Berkeland2002}%
  \BibitemOpen
  \bibfield  {author} {\bibinfo {author} {\bibfnamefont {D.~J.}\ \bibnamefont
  {Berkeland}}\ and\ \bibinfo {author} {\bibfnamefont {M.~G.}\ \bibnamefont
  {Boshier}},\ }\href {\doibase 10.1103/PhysRevA.65.033413} {\bibfield
  {journal} {\bibinfo  {journal} {Phys. Rev. A}\ }\textbf {\bibinfo {volume}
  {65}},\ \bibinfo {pages} {033413} (\bibinfo {year} {2002})}\BibitemShut
  {NoStop}%
\bibitem [{\citenamefont {Truppe}\ \emph
  {et~al.}(2017{\natexlab{b}})\citenamefont {Truppe}, \citenamefont {Williams},
  \citenamefont {Fitch}, \citenamefont {Hambach}, \citenamefont {Wall},
  \citenamefont {Hinds}, \citenamefont {Sauer},\ and\ \citenamefont
  {Tarbutt}}]{Truppe2017a}%
  \BibitemOpen
  \bibfield  {author} {\bibinfo {author} {\bibfnamefont {S.}~\bibnamefont
  {Truppe}}, \bibinfo {author} {\bibfnamefont {H.~J.}\ \bibnamefont
  {Williams}}, \bibinfo {author} {\bibfnamefont {N.~J.}\ \bibnamefont {Fitch}},
  \bibinfo {author} {\bibfnamefont {M.}~\bibnamefont {Hambach}}, \bibinfo
  {author} {\bibfnamefont {T.~E.}\ \bibnamefont {Wall}}, \bibinfo {author}
  {\bibfnamefont {E.~A.}\ \bibnamefont {Hinds}}, \bibinfo {author}
  {\bibfnamefont {B.~E.}\ \bibnamefont {Sauer}}, \ and\ \bibinfo {author}
  {\bibfnamefont {M.~R.}\ \bibnamefont {Tarbutt}},\ }\href
  {http://stacks.iop.org/1367-2630/19/i=2/a=022001} {\bibfield  {journal}
  {\bibinfo  {journal} {New Journal of Physics}\ }\textbf {\bibinfo {volume}
  {19}},\ \bibinfo {pages} {022001} (\bibinfo {year}
  {2017}{\natexlab{b}})}\BibitemShut {NoStop}%
\bibitem [{\citenamefont {Zhelyazkova}\ \emph {et~al.}(2014)\citenamefont
  {Zhelyazkova}, \citenamefont {Cournol}, \citenamefont {Wall}, \citenamefont
  {Matsushima}, \citenamefont {Hudson}, \citenamefont {Hinds}, \citenamefont
  {Tarbutt},\ and\ \citenamefont {Sauer}}]{Zhelyazkova2014}%
  \BibitemOpen
  \bibfield  {author} {\bibinfo {author} {\bibfnamefont {V.}~\bibnamefont
  {Zhelyazkova}}, \bibinfo {author} {\bibfnamefont {A.}~\bibnamefont
  {Cournol}}, \bibinfo {author} {\bibfnamefont {T.~E.}\ \bibnamefont {Wall}},
  \bibinfo {author} {\bibfnamefont {A.}~\bibnamefont {Matsushima}}, \bibinfo
  {author} {\bibfnamefont {J.~J.}\ \bibnamefont {Hudson}}, \bibinfo {author}
  {\bibfnamefont {E.~A.}\ \bibnamefont {Hinds}}, \bibinfo {author}
  {\bibfnamefont {M.~R.}\ \bibnamefont {Tarbutt}}, \ and\ \bibinfo {author}
  {\bibfnamefont {B.~E.}\ \bibnamefont {Sauer}},\ }\href {\doibase
  10.1103/PhysRevA.89.053416} {\bibfield  {journal} {\bibinfo  {journal} {Phys.
  Rev. A}\ }\textbf {\bibinfo {volume} {89}},\ \bibinfo {pages} {053416}
  (\bibinfo {year} {2014})}\BibitemShut {NoStop}%
\bibitem [{\citenamefont {Xie}\ \emph {et~al.}(2016)\citenamefont {Xie},
  \citenamefont {Bu},\ and\ \citenamefont {Yan}}]{Xie2016}%
  \BibitemOpen
  \bibfield  {author} {\bibinfo {author} {\bibfnamefont {D.}~\bibnamefont
  {Xie}}, \bibinfo {author} {\bibfnamefont {W.}~\bibnamefont {Bu}}, \ and\
  \bibinfo {author} {\bibfnamefont {B.}~\bibnamefont {Yan}},\ }\href
  {http://stacks.iop.org/1674-1056/25/i=5/a=053701} {\bibfield  {journal}
  {\bibinfo  {journal} {Chin. Phys. B}\ }\textbf {\bibinfo {volume} {25}},\
  \bibinfo {pages} {053701} (\bibinfo {year} {2016})}\BibitemShut {NoStop}%
\bibitem [{\citenamefont {Metcalf}\ and\ \citenamefont
  {Straten}(1999)}]{Metcalf1999}%
  \BibitemOpen
  \bibfield  {author} {\bibinfo {author} {\bibfnamefont {H.~J.}\ \bibnamefont
  {Metcalf}}\ and\ \bibinfo {author} {\bibfnamefont {P.}~\bibnamefont
  {Straten}},\ }\href@noop {} {\emph {\bibinfo {title} {Laser cooling and
  trapping}}}\ (\bibinfo  {publisher} {Springer},\ \bibinfo {year}
  {1999})\BibitemShut {NoStop}%
\bibitem [{\citenamefont {Tarbutt}(2015)}]{Tarbutt2015}%
  \BibitemOpen
  \bibfield  {author} {\bibinfo {author} {\bibfnamefont {M.~R.}\ \bibnamefont
  {Tarbutt}},\ }\href {http://stacks.iop.org/1367-2630/17/i=1/a=015007}
  {\bibfield  {journal} {\bibinfo  {journal} {New J. Phys.}\ }\textbf {\bibinfo
  {volume} {17}},\ \bibinfo {pages} {015007} (\bibinfo {year}
  {2015})}\BibitemShut {NoStop}%
\end{thebibliography}%

\end{document}